\documentclass[twocolumn,preprintnumbers,amsmath,amssymb,floatfix,superscriptaddress,nofootinbib]{revtex4-2}
\usepackage{amsmath,hyperref,amssymb}
\usepackage{graphicx}% Include figure files
\usepackage{dcolumn}% Align table columns on decimal point
\usepackage{bm}% bold math
\usepackage{verbatim}
\usepackage{tabularx}
\usepackage{slashed}
\usepackage{cancel}
\usepackage{hyperref}

\newcommand\eea{\end{eqnarray}}
\newcommand\bea{\begin{eqnarray}}

\def\beq{\begin{equation}}
\def\eeq{\end{equation}}

\newcommand{\be}{\begin{equation}}
\newcommand{\ee}{\end{equation}}
\newcommand{\ba}{\begin{align}}
\newcommand{\ea}{\end{align}}
\newcommand{\bg}{\begin{gather}}
\newcommand{\eg}{\end{gather}}
\newcommand{\bseq}{\begin{subequations}}
\newcommand{\eseq}{\end{subequations}}

\newcommand{\tr}{{\rm tr}}

\begin{document}
\title{The entropic $g$-theorem in general spacetime dimension}
\author{Horacio Casini}
\affiliation{Centro At\'omico Bariloche and CONICET, S.C. de Bariloche, R\'io Negro, R8402AGP, Argentina}
\author{Ignacio Salazar Landea}
\affiliation{Instituto de F\'\i sica de La Plata - CONICET, C.C. 67, 1900 La Plata, Argentina}
\author{Gonzalo Torroba}
\affiliation{Centro At\'omico Bariloche and CONICET, S.C. de Bariloche, R\'io Negro, R8402AGP, Argentina}
\date{\today}
\begin{abstract}
We establish the irreversibility of renormalization group flows on a pointlike defect inserted in a $d$-dimensional Lorentzian conformal field theory. We identify the impurity entropy $g$ with the quantum relative entropy in two equivalent ways. One involves a null deformation of the Cauchy surface, and the other is given in terms of a local quench protocol. Positivity and monotonicity of the relative entropy imply that $g$ decreases monotonically along renormalization group flows, and provides a clear information-theoretic meaning for this irreversibility.
\end{abstract}

\maketitle

%%%%%%%%%%%%%%%%%%%%%%%%%%%%%%%%%%%%%%%%%%%%%%%%%%%%%%%%%%%%%%%%%
%%%%%%%%%%%%%%%%%%%%%%%%%%%%%%%%%%%%%%%%%%%%%%%%%%%%%%%%%%%%%%%%%
%%%%%%%%%%%%%%%%%%%%%%%%%%%%%%%%%%%%%%%%%%%%%%%%%%%%%%%%%%%%%%%%%
%%%%%%%%%%%%%%%%%%%%%%%%%%%%%%%%%%%%%%%%%%%%%%%%%%%%%%%%%%%%%%%%%
\section{Introduction}\label{sec:intro}

Quantum information theory has emerged as a central tool for understanding quantum field theory. While a lot of progress has been made in Lorentz invariant systems using such methods \cite{Casini:2004bw, Casini:2012ei, casini2017modular, casini2017markov}, much less is known for non-Lorentz invariant theories. Investigating their dynamics is crucial, given that they are relevant in many areas of physics, including condensed matter and high energy physics, statistical mechanics and astrophysics. In this work we analyze a class of such theories: a pointlike defect immersed in an arbitrary dimensional conformal bulk (CFT). These defects can appear in gauge theories, Kondo models, impurities in quantum materials, etc. By using quantum information methods we will establish the irreversibility of defect renormalization group flows (RG) and we will provide a distinguishability measure for defects.

The main tool we will use is the quantum relative entropy, in particular its positivity and monotonicity properties. Computing the relative entropy in continuum field theories is usually problematic due to the appearance of infinities and an important step in our proof will be to construct a setup where the relative entropy is finite. Given this, we will use it to define an intrinsic distinguishability measure $g$ for the defect, with the property that $g$ decreases monotonically under defect RG flows. As a result, a consistent defect RG flow that connects UV and IR fixed points must obey $g_{UV}> g_{IR}$.

Our work is an extension (to higher dimensions) of the method we developed in \cite{Casini:2016fgb} for proving the $g$-theorem in $1+1$-dimensional bulk CFTs. In this dimensionality, the quantity $g$ is the entanglement entropy in the presence of the impurity, and its decrease was established by relating it to the relative entropy. For higher dimensional bulks, however, additional contributions to the entanglement entropy make it non-monotonic \cite{Kobayashi:2018lil} (see also \cite{Jensen:2018rxu}). Instead, $g$ has to be constructed directly from the relative entropy, and this is what we will do.

The information-theoretic results we will obtain apply to zero temperature real time field theories. They complement recent progress using euclidean methods at finite temperature, starting from \cite{Friedan:2003yc}. Indeed, our analysis was strongly motivated by \cite{Kobayashi:2018lil}, who conjectured that the euclidean sphere partition function in the presence of the defect should give the appropriate monotonic quantity. Recently this conjecture was proved for euclidean circular defects by \cite{Cuomo:2021rkm}. While our approach is inherently Lorentzian and has no clear euclidean analog, we will nevertheless be able to establish a connection with these works at fixed points. The combination of Lorentzian and euclidean results offers then a comprehensive characterization of RG flows for linear defects. A sample of other works on RG flows for defects with different dimensionalities includes \cite{Affleck:1991tk, Takayanagi:2011zk, Jensen:2013lxa, Jensen:2015swa, Casini:2018nym, Wang:2020xkc, Wang:2021mdq, Shachar:2022fqk, Castiglioni:2022yes}.

%%%%%%%%%%%%%%%%%%%%%%%%%%%%%%%%%%%%%%%%%%%%%%%%%%%%%%%%%%%%%%%%%
%%%%%%%%%%%%%%%%%%%%%%%%%%%%%%%%%%%%%%%%%%%%%%%%%%%%%%%%%%%%%%%%%
%%%%%%%%%%%%%%%%%%%%%%%%%%%%%%%%%%%%%%%%%%%%%%%%%%%%%%%%%%%%%%%%%
%%%%%%%%%%%%%%%%%%%%%%%%%%%%%%%%%%%%%%%%%%%%%%%%%%%%%%%%%%%%%%%%%
\section{Renormalization group flows on pointlike defects}\label{sec:RG}

Let us begin with a brief summary of the setup.
We consider a $d$-dimensional unitary CFT, with space-time coordiantes $x^\mu$, $\mu=0, \ldots, d-1$, and a pointlike defect located at $x^i=0$ and extended along the time direction $x^0$. We assume that at short distances the theory is described by a conformal fixed point, with symmetry group $SO(2, 1) \times SO(d-1) \subset SO(2,d)$. This is the subgroup of the full conformal group that keeps the defect fixed. The vacuum state for this theory is denoted by $\sigma$. We work in Minkowski signature $(-+\ldots+)$.

Now we introduce a relevant perturbation on the line defect,
\be\label{eq:RG1}
S= S_{DCFT}+ \int_{-\infty}^{\infty} dt\,\lambda\, \mathcal O
\ee
where $\mathcal O$ is a primary scalar operator of conformal dimension $\Delta$, and $\Delta<1$ so that the perturbation is relevant. Additional relevant interactions on the impurity can be added without affecting our analysis. The relevant interaction triggers a nontrivial RG flow on the defect, while the bulk remains conformal. (In contrast, relevant deformations in the bulk can lead to non-monotonic behavior on the defect; see e.g. \cite{Green:2007wr}). We assume that at long distances the theory flows to an IR fixed point. The vacuum state of (\ref{eq:RG1}) is denoted by $\rho$.

A key operator for characterizing the dynamics of the theory is the energy-momentum tensor. The impurity generically adds delta-function localized contributions, so that
\be
T^{\mu\nu}(x)= T_b^{\mu\nu}(x)+\delta^{d-1}(x) \delta^{\mu}_0 \delta^{\nu}_0 \theta(x)\,.
\ee
At a fixed point, conservation and conformal invariance imply $\langle \theta \rangle =0$ and
\bea\label{eq:aT}
\langle T_{00}\rangle &=&- \frac{d-2}{d}\frac{a_T}{r^d}\;,\;\langle T_{0i}\rangle =0\nonumber\\
 \langle T_{ij}\rangle &=& \frac{a_T}{r^d} \left(\frac{x_i x_j}{r^2}- \frac{2}{d}\delta_{ij}\right)\,,
\eea
with $a_T$ a constant. Note that $\langle T_{\mu\nu} \rangle=0$ for $d=2$, and this is made explicit by the choice of normalization for $a_T$. On the other hand, the one-point function is nontrivial for $d>2$.

Away from the fixed point the one-point function receives corrections from the relevant deformation. As discussed in \cite{Casini:2018nym} and reviewed below, the stress tensor one point function is characterized in terms of a single function $h(r)$. We will use this structure for our definition of impurity entropy. At second order in perturbation theory we expect, on dimensional grounds, a correction
\be\label{eq:UV1}
\Delta \langle T_{\mu\nu}(r) \rangle \sim \frac{1}{r^d}\,\lambda^2 r^{2(1-\Delta)}\;,\; r \ll \lambda^{-1/(1-\Delta)}\,.
\ee
On the other hand, for large $r$ the defect theory becomes conformal again, and then
\be\label{eq:IR1}
\Delta \langle T_{\mu\nu}(r) \rangle   \sim \frac{a_T^{IR}-a_T^{UV}}{r^d}\;,\; r \gg \lambda^{-1/(1-\Delta)}\,.
\ee

%%%%%%%%%%%%%%%%%%%%%%%%%%%%%%%%%%%%%%%%%%%%%%%%%%%%%%%%%%%%%%%%%
%%%%%%%%%%%%%%%%%%%%%%%%%%%%%%%%%%%%%%%%%%%%%%%%%%%%%%%%%%%%%%%%%
%%%%%%%%%%%%%%%%%%%%%%%%%%%%%%%%%%%%%%%%%%%%%%%%%%%%%%%%%%%%%%%%%
%%%%%%%%%%%%%%%%%%%%%%%%%%%%%%%%%%%%%%%%%%%%%%%%%%%%%%%%%%%%%%%%%
\section{Defect entropy from quantum relative entropy}\label{sec:Srelg}

Motivated by previous works \cite{Casini:2016fgb, Casini:2016udt, Casini:2018nym}, our goal is to use the quantum relative entropy to compare the states $\rho$ and $\sigma$ above, in order to characterize nonperturbatively the impurity RG flow. To this aim, consider a spherical region of radius $R$ centered around the defect. Its domain of dependence $V$ is the causal diamond $|x^0|+r \le R$. The density matrices of the previous two states that we introduced, reduced to this region, are written as $\sigma_R$ and $\rho_R$, and the relative entropy is defined as
\be\label{eq:Srel1}
S_{rel}(\rho_R|\sigma_R)= \tr_V\left( \rho_R(\log \rho_R-\log \sigma_R)\right)\,.
\ee
This is a measure of distinguishability between the two density matrices.
Note that we are comparing the theory with and without relevant interaction; the two states act then on the same algebra of operators, and we can compute their relative entropy.  

The relative entropy can also be interpreted as a difference of free energies. Indeed, in terms of the modular Hamiltonian $H_\sigma=-\log \sigma_R$ for the fixed point theory, (\ref{eq:Srel1}) becomes
\be\label{eq:Srel2}
S_{rel}(\rho_R|\sigma_R) = \Delta \langle  H_\sigma \rangle - \Delta S\,.
\ee
Here the first energy term is the difference of the modular Hamiltonian expectation values for the two states, $\Delta \langle  H_\sigma \rangle=\tr\left((\rho-\sigma) H_\sigma\right)$. The modular Hamiltonian for the state $\sigma$ of the conformal defect is given by \cite{Casini:2018nym}
\be\label{eq:modham}
 H_\sigma= \int_\Sigma d^{d-1}x\,\eta^\mu\xi^\nu\, T_{\mu\nu}
\ee
with $\eta^\mu$ the unit normal to the Cauchy surface $\Sigma$, and 
\be\label{eq:killing}
\xi^\nu=\frac{\pi }{R} \left(R^2-(x^0)^2-(x^i)^2,-2 x^0 x^i \right)
\ee
is the Killing vector for a conformal transformation that keeps the sphere fixed \cite{Casini:2011kv}. On the other hand, the entropy term in (\ref{eq:Srel1}) is the difference in von Neumann entropies, 
\be
\Delta S=-\tr( \rho \log \rho)+\tr(\sigma \log \sigma)\,.
\ee

The relative entropy is positive and increases monotonically with $R$.
The RG flow is accessed by varying the radius $R$ compared to the typical RG distance scale, $\lambda^{-1/(1-\Delta)}$. Since the two states differ by the relevant deformation (\ref{eq:RG1}), we expect that in the limit $R \to 0$ they become indistinguishable in the sense that $S_{rel} \to 0$. The intuitive interpretation for monotonicity is that increasing $R$ allows to use more operators to distinguish the two states and hence $S_{rel}$ increases. 

These properties are what we need in order to establish the irreversibility of defect RG flows. However, the main obstacle is that the relative entropy is often divergent in continuum QFT. To see this, consider a spatial Cauchy surface at $x^0=0$. With $\eta_\mu=\delta_\mu^0$ and the one-point function (\ref{eq:aT}), we find that (\ref{eq:modham}) at a fixed point diverges as $\langle H_\sigma \rangle \sim a_T \frac{R}{\epsilon}$, with $\epsilon$ a short distance cutoff for the spatial integral. This leading divergence cancels out in $\Delta  \langle H_\sigma \rangle$; however, the leading perturbative correction (\ref{eq:UV1}) gives
\be\label{eq:problem}
\Delta  \langle H_\sigma \rangle \sim \left(\lambda^2 \epsilon^{1-2\Delta} \right) R\,.
\ee
Then the relative entropy diverges when $1/2 \le \Delta \le 1$ (the upper bound being, as before, the condition that the interaction is relevant). We then know that this term linear in $R$ should be positive and (trivially) monotonically increasing, but it does not provide an intrinsic quantity that we can associate to fixed points. We need to access subleading terms in the relative entropy, but we cannot just subtract the linear term (\ref{eq:problem}) -- this would ruin the information-theoretic interpretation and the monotonicity property.

\begin{figure}[ht]
\begin{center}
\includegraphics[scale=0.3]{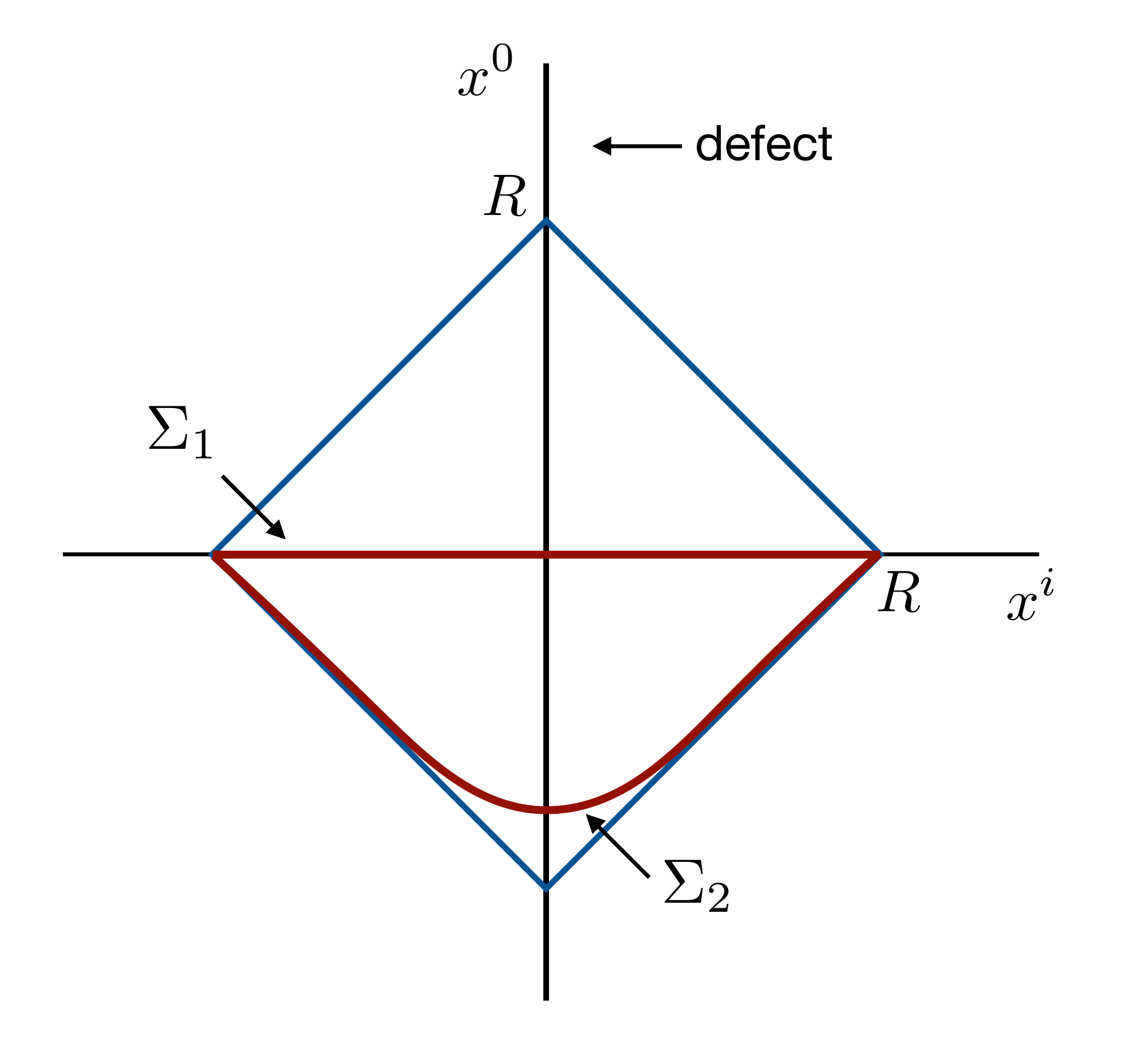}
\caption{Causal development for a sphere of radius $R$, with a line defect at the center, and two Cauchy slices $\Sigma_1$ (constant time) and $\Sigma_2$ (almost null). The relative entropy depends on the choice of Cauchy surface.} \label{fig:null}
\end{center}
\end{figure}

Instead, we will consider the relative entropy on a Cauchy surface that approaches the past line cone, centered at the defect.  A bi-dimensional cut is depicted in Fig. \ref{fig:null}, with the almost null surface being $\Sigma_2$. This generalizes the method of \cite{Casini:2016fgb} for the $g$-theorem in 1+1 dimensions. The basic idea is that $\Delta \langle H_\sigma \rangle$ and hence the relative entropy depend on the choice of Cauchy surface $\Sigma$. This dependence can in turn be used to eliminate the linear term (\ref{eq:problem}).

The dependence on $\Sigma$ is due to the fact that the two states $\sigma$ and $\rho$, though defined on the same algebra of operators, have different time evolution. As in \cite{Casini:2016udt} we can work in a conformal interaction picture where operators are in the Heisenberg picture of the UV defect CFT Hamiltonian, while $\rho$ evolves with the interaction Hamiltonian $H_{int}= \delta^{d-1}(x)\,\lambda \mathcal O(x)	$ (the state $\sigma$ is time-independent in this picture). Denote by $U_\Sigma$ the evolution operator for $H_{int}$ from $t \to -\infty$ up to the Cauchy surface $\Sigma$. 
Starting from the global vacuum of the UV fixed point, $\sigma=  | 0 \rangle \langle 0 |$, the state $\rho$ is obtained as $\rho= U_\Sigma \sigma U_\Sigma^\dag$, which exhibits explicitly the dependence of $\rho$ on $\Sigma$. After computing the reduced state $\rho_R$, this results in  $\Delta \langle H_\sigma \rangle$ depending on $\Sigma$. Equivalently, in a Heisenberg representation, the identification of the algebra of operators in both theories depends on the choice of Cauchy surface \cite{Casini:2016udt}.

\begin{figure}[ht]
\begin{center}
\includegraphics[scale=0.30]{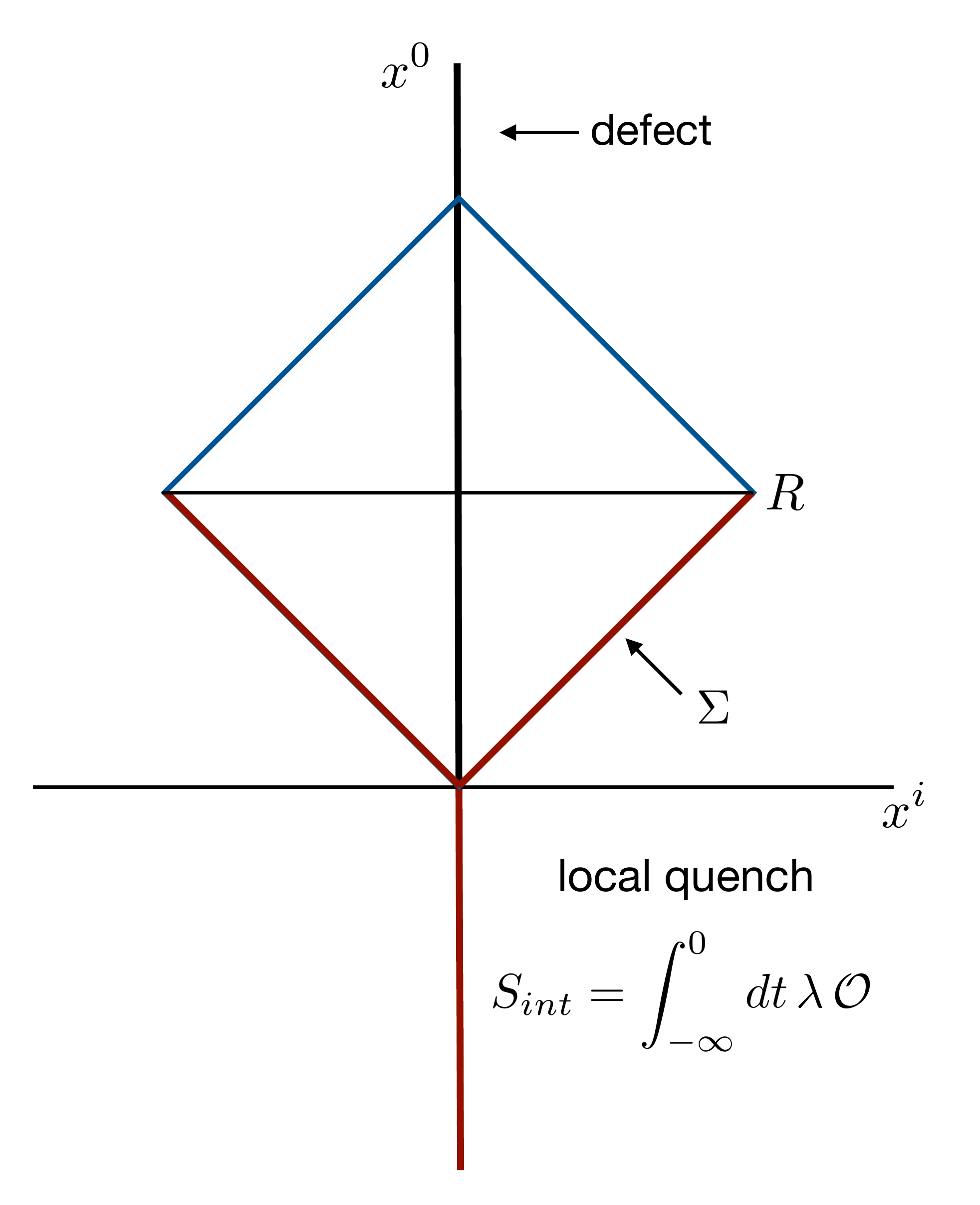}
\caption{The relevant interaction is turned off at $x^0=-R$, after which the evolution operator becomes conformal. The vacuum state at $x^0=-R$ obtained by euclidean evolution from $t \to -\infty$ defines a reduced density matrix $\rho_R$ on $\Sigma$.} \label{fig:quench}
\end{center}
\end{figure}

Another way to view this is via a local quench. The relevant interaction (\ref{eq:RG1}) is turned on between $-\infty < x^0 < -R$ and near $x^0 \approx -R$ we turn it off (in some smooth way). The global state $\rho$ is defined by the time evolution including the relevant interaction, while for $x^0>-R$ both states $\sigma$ and $\rho$ evolve with the same conformal Hamiltonian. We then compute the relative entropy for both states in the setup of Fig. \ref{fig:quench}. In this construction, the state $\rho$ for the relative entropy is naturally defined on the null Cauchy surface $\Sigma$ shown in the figure. But because after $x^0=-R$ both $\rho$ and $\sigma$ evolve with the same Hamiltonian, we can actually take other Cauchy surfaces in the same causal development, and the relative entropy will not change. Both setups are physically distinct, but the concrete calculations leading to the cancellation of (\ref{eq:problem}) are the same. We turn to these next.

Let us approach the null limit in terms of a family of spacelike hyperboloids of curvature $a$, with $a \to 0$
\be\label{eq:hyperboloid}
(x^0+\sqrt{R^2+a^2})^2-r^2=a^2\;,\;0\le r \le R\;,\;x^0<0\;.
\ee
In this case, the nonvanishing component of the Killing vector near the defect is $\xi^0= 2\pi a + O(a^2)$, which should be contrasted with $\xi^0 = 2\pi R$ for the constant-time surface. Repeating the previous calculation for the change in modular Hamiltonian for the almost null Cauchy surface then gives a $\Delta \langle H_\sigma \rangle$ that is independent of $R$. This constant can be simply subtracted from the relative entropy without altering the monotonicity property $\partial_R S_{rel}(\rho_R|\sigma_R) \ge 0$. It is also possible to define a regularization scheme where the constant is set to zero.

Intuitively, the choice $x^0=0$ (namely $\Sigma_1$ in the figure) places the defect in a region with modular temperature $T \sim 1/R$, and this is the origin of (\ref{eq:problem}). Instead, taking a Cauchy surface that approaches the null limit gives $T \to \infty$ near the defect. This high temperature limit eliminates the diverging distinguishability of the two states.

In the setup of a null Cauchy surface, let us define the difference of impurity $g$-functions as
\be\label{eq:proposal}
S_{rel}(\rho_R|\sigma_R)= - \log \frac{g(R)}{g(0)}\,.
\ee
At the end of the following analysis we will arrive to a natural definition of $g$ at fixed points (see (\ref{eq:g-fp}) below) and so (\ref{eq:proposal}) determines $g(R)$.
We need to show that when $R \to \infty$ the relative entropy decomposes as a difference $- \log (g_{IR}/g_{UV})$, where $g_{UV}$ and $g_{IR}$ are associated to the UV and IR fixed points, respectively. The entropy contribution $\Delta S$ in (\ref{eq:Srel2}) already gives a difference between fixed point quantities, and it remains to establish the same for $\Delta \langle H_\sigma \rangle$.

The modular Hamiltonian can be written as an integral over the future light cone emanating from the past tip of the causal diamond. For convenience we set the past tip of the double cone at the origin, and use the coordinate $\lambda=(x^0+r)/2$, that is equal to both $x^0$ and $r$ on the light cone. We have  
\be
H_\sigma=\frac{2\pi}{R}\,\int d^{d-2}\Omega\,\,\,\int_0^{R} d\lambda \,  \lambda^{d-1}\, \, (R-\lambda)\,\, T_{\lambda\lambda}(\lambda,\Omega)\,,
\ee
with $\Omega$ describing the angular coordinates over the unit sphere and 
\be
T_{\lambda\lambda}= T_{00}+ \hat x^i\,\hat x^j T_{ij}\,.
\ee

The traceless and conserved bulk stress-tensor one point function in presence of the defect can be parametrized as \cite{Casini:2018nym}
\bea
 \langle T_{00}  \rangle &=& -h(r)\;,\hspace{1cm} \langle T_{0i}  \rangle=0\\
 \langle T_{ij}  \rangle &=&- \frac{1}{d-1}\,h(r)\, \delta_{ij}+f(r) \left(\hat x_i \hat x_j- \frac{\delta_{ij}}{d-1} \right)\,.\nonumber
\eea
The conservation condition gives a differential equation 
%\be
%f'(r)+ \frac{d-1}{r} f(r) - \frac{1}{d-2}h'(r) =0\,,
%\ee
with solution
\be
f(r) = -\frac{1}{d-2} \frac{1}{r^{d-1}} \int_r^\infty du\,u^{d-1} h'(u)\,.
\ee
At fixed points we have 
\be\label{eq:larger}
h(r) = \frac{(d-2)\, a_T}{d}\, \frac{1}{r^d}\,,\,
f(r) = a_T\, \frac{1}{r^d}\,.
\ee
The null component of the stress tensor is 
\be
\langle T_{\lambda\lambda}\rangle =\frac{1}{(d-1)} \, \left(- d\, h(r)+ (d-2)\, f(r)\right)\,.
\ee
Using (\ref{eq:larger}) this vanishes for a conformal defect.

Let us now compute the change in the modular Hamiltonian
\bea
\Delta \langle H_\sigma\rangle &=&-\frac{2\pi{\rm Vol}(S^{d-2})}{(d-1)R}\int_0^R d\lambda (R-\lambda)\\
&&\left[d\, \lambda^{d-1}\Delta h(\lambda)+ \int_\lambda^\infty du \,u^{d-1}\Delta h'(u)  \right]\nonumber\,.
\eea
Integrating by parts the $h'(u)$ factor, changing the order of integration in the double integral, and performing the integration over $\lambda$, one obtains
\bea
\Delta \langle H_\sigma\rangle &=&\pi\frac{{\rm Vol}(S^{d-2})}{R}\,\Big[\int_0^R d\lambda\,\lambda^d \Delta h(\lambda) \nonumber\\
&&+ R^2 \int_R^\infty d\lambda\,\lambda^{d-2}\Delta h(\lambda) \Big]\,.
\eea
At large $R$, the integral is dominated by the large $ \lambda$ behavior (\ref{eq:larger}), which gives
\be
\lim_{R \to \infty}\Delta \langle H_\sigma\rangle =- \frac{2\pi(d-2)}{d}\,{\rm Vol}(S^{d-2})\,(a_T^{IR}- a_T^{UV})\,.
\ee
This is a difference of fixed point quantities.

We conclude that at large $R$, the relative entropy on the Cauchy surface that approaches the past or future lightcone for the sphere of radius $R$ indeed becomes a difference of fixed point quantities, 
\bea\label{eq:Sinf1}
\lim_{R \to \infty} S_{rel}(\rho_R|\sigma_R)& =&\frac{2\pi(d-2)}{d}\,{\rm Vol}(S^{d-2})\,(a_T^{UV}- a_T^{IR})\nonumber\\
&+& \lim_{R \to \infty} \left(S(\sigma_R)-S(\rho_R) \right)\,.\; \qquad  \quad
\eea
This defines the difference of defect $g$-entropies
\be\label{eq:gIRgUV}
\lim_{R \to \infty} S_{rel}(\rho_R|\sigma_R) = - \log \frac{g_{IR}}{g_{UV}}
\ee
with $g_{UV}=g(R=0)$, $g_{IR}= g(R \to \infty)$. 

Given (\ref{eq:gIRgUV}), we can identify a fixed point quantity $\log g$ from the first and third terms (resp. second and forth terms) in (\ref{eq:Sinf1}),
\be\label{eq:g-fp}
\log g =\frac{2\pi(d-2)}{d}{\rm Vol}(S^{d-2})\,a_T+ S(\sigma)-S_{bulk}(\sigma)
\ee
where the conformal bulk contributions to the entanglement entropy have been subtracted (they cancel out in the relative entropy). The running $g$-function $g(R)$ is then determined in terms of the relative entropy (\ref{eq:proposal}) and the fixed point value (\ref{eq:g-fp}).

%%%%%%%%%%%%%%%%%%%%%%
%%%%%%%%%%%%%%%%%%%%%%
%%%%%%%%%%%%%%%%%%%%%%
\section{Entropic $g$-theorem}\label{sec:gthm}

Having demonstrated that at large $R$ the relative entropy splits as a difference of fixed-point intrinsic quantities, we are ready to establish the irreversibility of defect RG flows.

First, positivity of the relative entropy implies that
\be\label{eq:Sinf2}
\log g_{UV} - \log g_{IR} >0\,.
\ee
Furthermore, the relative entropy at finite $R$ provides a natural definition for a running $g$-function, given above in (\ref{eq:proposal}). Monotonicity of the relative entropy, $S_{rel}'(R) \ge 0$ then implies the monotonic decrease
\be\label{eq:gmon}
g'(R) \le 0\,.
\ee
Eqs. (\ref{eq:Sinf1}) -- (\ref{eq:gmon}) are the main results of the paper. They establish the entropic $g$-theorem in general space-time dimension, and the irreversibility has a clear interpretation in terms of the increase in distinguishability as measured by the relative entropy.

To end, let us compare our results with those obtained in euclidean field theory. We note that our proposal (\ref{eq:g-fp}) agrees with the euclidean sphere partition function computed in \cite{Kobayashi:2018lil,Lewkowycz:2013laa}. Therefore, the fixed point statement (\ref{eq:Sinf2}) agrees with the recent proof in \cite{Cuomo:2021rkm}. The euclidean treatment, however, is more analogous to our calculation of the expectation value of the modular Hamiltonian at $x^0=0$. As we have seen, this requires subtracting a term linear in $R$, and the interpretation as a relative entropy is lost.

%\bigskip

%%%%%%%%%%%%%%%%%%%%%%
%%%%%%%%%%%%%%%%%%%%%%
%%%%%%%%%%%%%%%%%%%%%%
\section{Conclusions}\label{sec:concl}

In this work we have established the irreversibility of RG flows on defects embedded in $d$-dimensional Lorentzian CFTs. We achieved this in terms of the quantum relative entropy for the reduced vacuum density matrices with and without the relevant deformations. We argued that for a Cauchy surface that approaches the past lightcone of a sphere of radius $R$, the relative entropy is free of UV divergences and leads to a notion of impurity $g$-function at fixed points. Positivity of relative entropy implies $\log g_{UV} > \log g_{IR}$, and its monotonicity allows to construct an interpolating function $g(R)$ with $g'(R) <0$ and $g(0)= g_{UV}, g(\infty) = g_{IR}$. From this information-theoretic perspective, the irreversibility is due to the increase in distinguishability between the two states.

At fixed points, we have shown that the quantity $g$ obtained from the relative entropy agrees with a regularized version of the defect free energy on the sphere \cite{Kobayashi:2018lil}. Our result $\log g_{UV} > \log g_{IR}$ is then equivalent to \citep{Cuomo:2021rkm}. Physically, however, the information-theoretic methods we have employed do not have a clear euclidean counterpart. It would be interesting to explore their connection further, perhaps by developing a sum rule for pointlike defects (see \cite{Casini:2018nym}) that could be compared to the gradient formula of \cite{Friedan:2003yc, Cuomo:2021rkm}. Finally,
a main direction of future work is to extend the present methods to defects of other dimensionalities. This is under current investigation.

\section*{Acknowledgments} 

HC and GT are supported by
CONICET (PIP grant 11220200101008CO), ANPCyT (PICT 2018-2517), CNEA, and Instituto Balseiro, Universidad Nacional de Cuyo. HC acknowledges an ``It From Qubit" grant of the Simons Foundation. ISL would like to thank IB and ICTP for hospitality.

\bibliography{EE}{}
\bibliographystyle{utphys}

\end{document}